\input ppltexa.sty
\input epsf
\mag\magstephalf
\hsize 130 truemm 
\hoffset=17.55mm
\newdimen\vu \vu=12 truebp
\parskip 0pt plus .1\vu \parindent 1.5\vu 
\def\section#1\par{\vskip 1.5\vu plus .3\vu
\vskip 0in plus 1in\penalty -50\vskip 0in plus -1in
\noindent{\sc #1}\par\penalty 1000\noindent}
\def\figinsert#1#2{\iffalse \vskip #2truemm plus 5truemm \else
\epsfysize=#2truemm
\advance\epsfysize by -10truemm \vskip 0truemm plus 2.5truemm
\advance\epsfysize by -\baselineskip
\centerline{\epsfbox{f#1.ps}}\vskip 5truemm plus 2.5truemm
\fi \ignorespaces}
\let\pn=\cal
\def\Dscr{{\pn D}}\def\Cscr{{\pn C}} 
\baselineskip 1.1\vu
\centerline{{\uppercase{\bf
Long-Time Correlations in Stochastic Systems}}\footnote*{%
Presented at the US-Japan Workshop on
Statistical Physics and Chaos in Fusion Plasmas,
Austin, Texas, December 13--17, 1982.
Published in {\it Statistical Physics and Chaos in Fusion Plasmas},
edited by C. W. Horton and L. E. Reichl,
Wiley, New York, 1984, pages 33--42.}}
\vskip 1.5\vu plus .3\vu minus 1pt
\centerline{Charles F. F. Karney}
\centerline{Plasma Physics Laboratory, Princeton University,}
\centerline{Princeton, New Jersey 08544}
\baselineskip 1.1\vu

\section Abstract

In recent years, there has been considerable
interest in understanding the motion in Hamiltonian systems when phase
space is divided into stochastic and integrable regions.  This paper
studies one aspect of this problem, namely, the motion of trajectories
in the stochastic sea when there is a small island present.
The results show that the particle
can be stuck close to the island for very long times.  For the standard
mapping, where accelerator modes are possible, it appears that the mean
squared displacement of particles in the stochastic sea may increase
faster than linearly with time indicating non-diffusive behavior.

\section Introduction

Many important problems in physics are described by Hamiltonians of two
degrees of freedom.  Examples are the motion of a charged particle in
electro\-static waves, the motion of a charged particle in various
magnetic confinement devices, the acceleration of a particle bouncing
between a fixed and an oscillating wall, the wandering of magnetic
field lines, etc.  In such systems, there is usually a range of
parameters (normally when the coupling between the two degrees of
freedom is large), where the motion in nearly the whole of phase space
is stochastic.  Such behavior is seen for instance in the
standard mapping (Chirikov, 1979),
$$r_t-r_{t-1}=-k\sin\theta_{t-1},
\qquad \theta_t-\theta_{t-1}=r_t.$$
When $k$ is large most of phase space is stochastic.  However, there may
still be small islands present; stochastic trajectories can wander
close to these islands and remain there for a long time leading to
unexpectedly long correlations.  The effect of these correlations can be
dramatic.  The simplest approximation for the diffusion coefficient,
$$\Dscr=\lim_{t\to\infty}{\ave{(r_t-r_0)^2}\over2t}$$
(where the average is over some appropriate ensemble), is given by
assuming that the phase $\theta$ is a random variable in the equation for $r$.
This gives the ``quasi-linear'' result $\Dscr=\Dscr_{\subrm{ql}}=\quarter
k^2$.  However, a numerical determination (Karney {\it et al.}, 1982) of
the diffusion at $k=6.6$,
where the random phase approximation might be expected to be accurate,
gave $\Dscr/\Dscr_{\subrm{ql}} \sim 80$.
At this
value of $k$ there is an island (``accelerator mode'') present in the
stochastic sea.  This leads to long-time correlations in the
acceleration of the particle and an enhanced diffusion coefficient.

In this paper, we examine more closely the effect these islands
have on a stochastic trajectory.  As far as determining the effect on
the correlation function, this involves determining how ``sticky'' the
island is.  Given that the stochastic trajectory comes within a certain
distance of the boundary of the island, how long do we expect it to stay close to the
island?  This approach is inspired by work of Channon and
Lebowitz (1981) on the correlations of a trajectory in the
stochastic band trapped between two KAM surfaces in the H\'enon map.  
Similar work has been carried out on the whisker map by Chirikov
and Shepelyansky (1981); this work is being extended by B. V.
Chirikov and F. Vivaldi.  The work reported herein is described in more
detail by Karney (1983).

\section  Derivation of Mapping

Far into the stochastic regime for a general mapping,
the islands which appear via tangent
bifurcations are very small and exist only for a small interval in
parameter space.  This allows us to approximate them by
a Taylor expansion in both
phase and parameter space about the tangent bifurcation point
retaining only the leading terms.  This was carried out by Karney {\it et
al.}\  (1982)
where the resulting mapping was reduced to a canonical form
$$Q:\qquad y_t-y_{t-1} = 2(x_{t-1}^2-K),
\qquad x_t-x_{t-1}=y_t.$$
Here $K$ is proportional to $k-k_{\subrm{tang}}$ ($k_{\subrm{tang}}$ is
the parameter value where the tangent bifurcation takes place) and $x$
and $y$ are related to the original phase space coordinates by a smooth
transformation.  The quadratic mapping $Q$ represents an approximation of the
general mapping close to the point of tangent bifurcation.
For $0<K<1$, this mapping has stable (elliptic) and unstable
(hyperbolic) fixed points at $(x,y)=(\mp \sqrt K,0)$, respectively.  The
elliptic fixed point is usually surrounded by integrable trajectories
(KAM curves) which define a stable region (the
island) in which the motion is bounded.  An
example of island structure is shown in Fig.\ 1 for
$K=0.1$ (the value of $K$ at which extensive numerical calculations have
been carried out).
\par\topinsert\figinsert{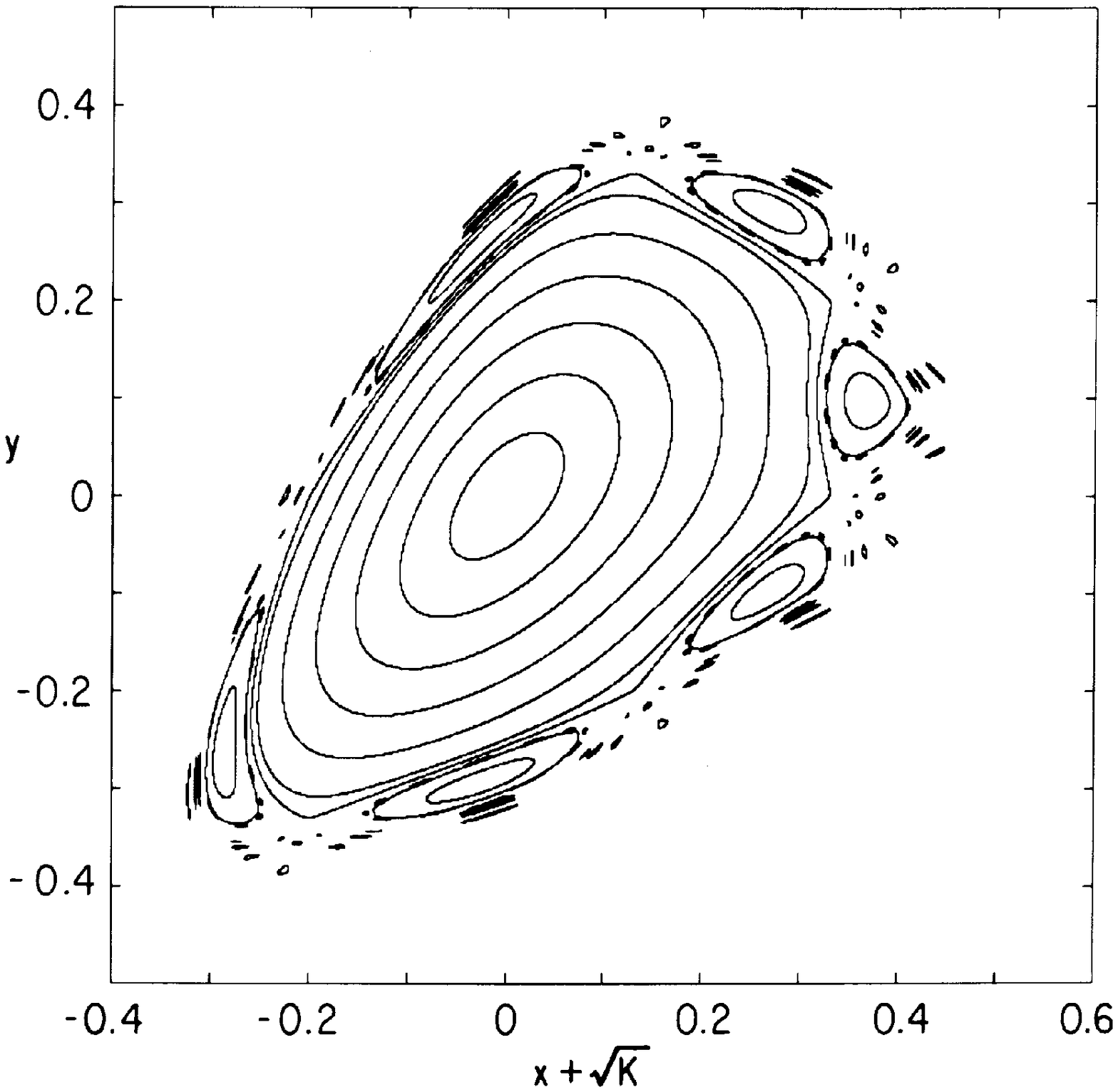}{85}
{\baselineskip 1\vu
\centerline {Fig.\ 1.  Some islands of the quadratic map $Q$
for $K=0.1$.}}\vskip 2truemm plus 1truemm\endinsert

Referring to the islands shown in Fig.\ 1, consider a
particle which at $t=0$ is close to, but outside, the islands.
Initially,
the particle will stay close to the islands; however as we let
$t\to\pm\infty$, we find $(x,y)\to(\infty,\pm\infty)$.  It is just such
trajectories we are interested in, because they correspond to particles
in the stochastic region of the general mapping approaching the islands,
staying there for some time (and contributing to long-time
correlations), and then escaping back to the main part of the stochastic
region.

\def\xmax{x_{\subrm{max}}}\def\xmin{x_{\subrm{min}}}\def\qstar{Q^\ast}%
What we need is some way of bringing these particles back to the
vicinity of the island.   We do this by defining an $L\times L$ square
around the island.  This square spans the region $\xmin\le x<\xmax$ and
$-\half L\le y<\half L$ where $L=\xmax-\xmin$.  Whenever an orbit
leaves this square at $(x_t,y_t)$, we pick a new initial condition
$$(x_0,y_0)=(x_t-mL,\,y_t-nL)$$
with $m$ and $n$ being integers chosen so that $(x_0,y_0)$ lies inside
the square.  We also record the length $t$ of the previous orbit
segment.  This procedure defines the periodic quadratic map
$\qstar$, which can be
shown to sample the orbits close to the island
in the same way that the general mapping does.
Examples of the orbits of $\qstar$ are shown in Fig.\ 2 for the same
parameters as for Fig.\ 1.
\par\topinsert\figinsert{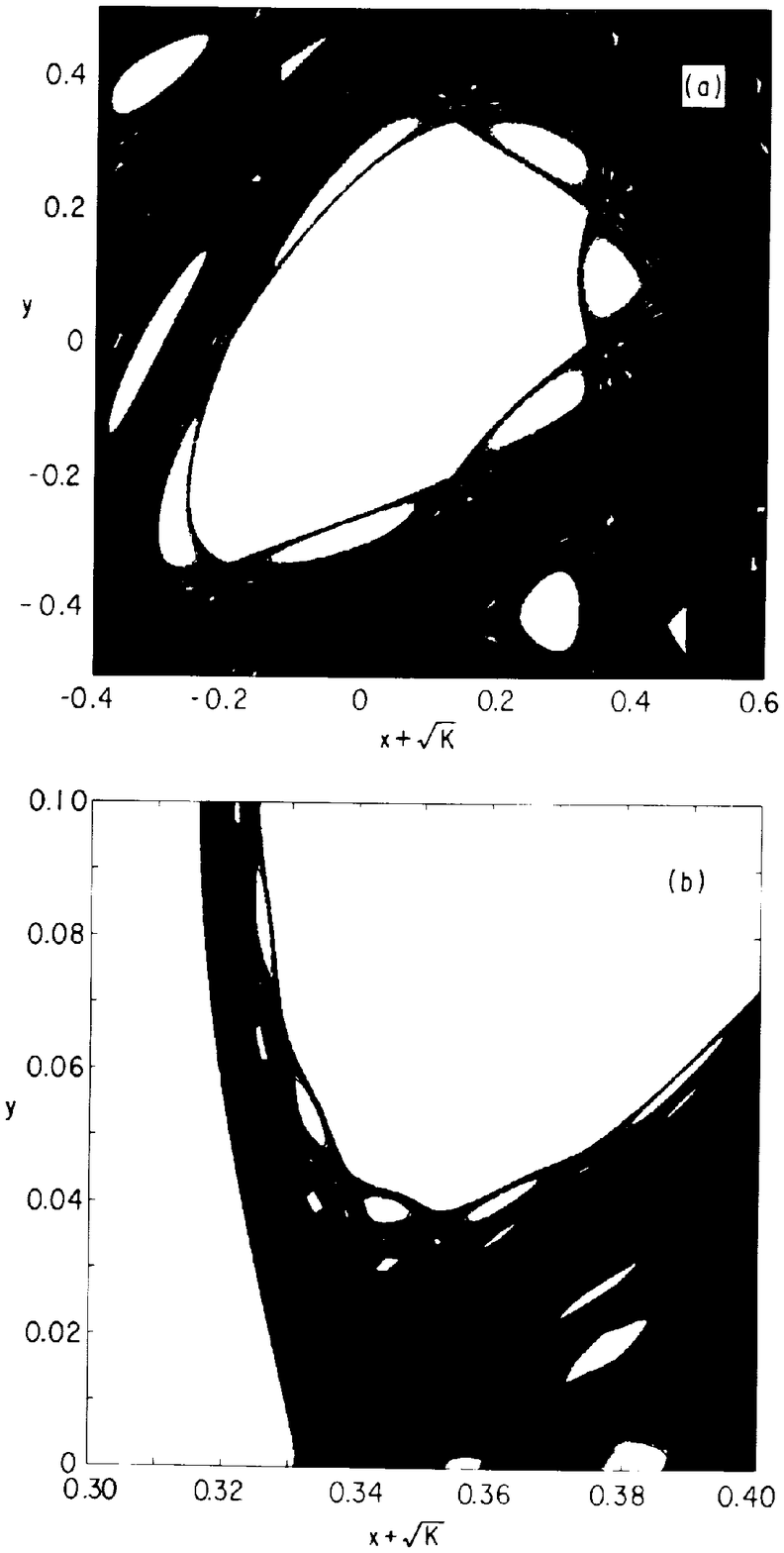}{161}
{\baselineskip 1\vu
\noindent Fig.\ 2.  (a) Stochastic trajectories for periodic quadratic map
$\qstar$ for $K=0.1$.  (b) An enlargement of a portion of (a).
Here $\xmin+\sqrt K=-0.4$, $\xmax+\sqrt K=0.6$.\par}
\vskip 5truemm plus 1truemm\endinsert

One useful way of looking at $\qstar$ is as a magnification of a small
region near a tangent bifurcation in the general mapping.  The
difference is that once the trajectory leaves the vicinity of the
islands, it is immediately re-injected on the other side of the
islands.  In the general map, the trajectory will spend some long time,
which depends on the ratio of the size of the islands to the total
accessible portion of phase space, in the stochastic sea before coming
back to the vicinity of the islands.

Assuming that the long-time behavior of stochastic orbits is dominated
by the region close to the islands, there are two advantages to
reducing the problem to a study of $\qstar$.  Firstly, since $\qstar$
describes the behavior of most islands far into the stochastic regime,
the properties of many mappings may be treated by looking at a special
mapping $\qstar$ which depends only on a single parameter $K$.  The
second advantage is that the properties of orbits close to the islands
may be studied much more efficiently because there is no need to follow
orbits while they spend a long and uninteresting time far from the
islands.

\section Trapping Statistics

The prescription for numerically determining the stickiness of the island
system in $Q$ is to compute a long orbit in the stochastic region of
$\qstar$.  The orbit
is divided into segments at those points where it leaves the $L\times L$
square.  The main results of the
calculation are then the {\it trapping statistics} $f_t$ which are
proportional to the number of orbit segments which have a
length of $t$.
Suppose that the total length of the orbit is $T$ and $N_t$ is the
number of segments of length $t$.  If $T$ is so large that we can ignore
partial segments at the ends of the orbit,
then we have $\sum tN_t = T$; the total number of segments is
$N=\sum N_t$.  The trapping statistics are defined by $f_t=N_t/T$ and
are therefore normalized so that $\sum tf_t=1$.  The mean length of the
orbits is given by $\alpha= 1/\sum f_t$ ($= T/N$).  The probability
that a particular segment has length $t$ is $p_t=\alpha f_t$
($=N_t/N$).  If an arbitrary point is chosen in the orbit, then $tf_t$
is the probability that this point belongs to a segment of length $t$
and $f_t$ is the probability that it belongs to the beginning, say, of
a segment of length@$t$.

The survival probability
$$P_t=\sum_{\tau=t+1}^\infty p_\tau$$
is the probability that an orbit beginning in a segment at $t=0$
is still trapped in
the same segment at time $t$.  Note that $P_0=1$ as required.  This is
the quantity studied by Channon and Lebowitz (1980) and Chirikov and Shepelyansky (1981).
The correlation function
$$C_\tau=\sum_{t=\tau}^\infty (t-\tau)f_t
=\sum_{t=\tau}^\infty P_t/\alpha$$
is the probability that a particle is trapped in the same segment at
two times $\tau$ apart.  Again, we have $C_0=1$.

There is another way of interpreting $C_\tau$:
Consider a drunkard who executes a one-dimensional random walk with
velocity $v=dr/dt=\pm1$.  The direction of each step is chosen randomly,
while the durations of the steps are chosen to be the lengths of
consecutive trapped segments of $\qstar$.  Then for integer $\tau$,
$C_\tau$ is just the usual correlation function for such a process,
i.e., $\ave{v_tv_{t+\tau}}_t$.  The behavior of this random-walk process
is similar to the behavior of an orbit in the general mapping when
two accelerator modes with opposite values of the acceleration are
present.  (This is the case with the first-order accelerator modes for
the standard mapping.)

A diffusion coefficient may be defined by
$$D=\half C_0+\sum_{\tau=1}^\infty C_\tau=\sum \half t^2 f_t.$$
This gives the diffusion rate for the drunkard in the random-walk
problem above.  It is also related to the diffusion coefficient for
the general mapping.

\section Results for \uppercase{$K=0.1$}

We have measured $f_t$ for $K$ between 0 and 1.3 at intervals of 0.05,
and at most of the values of $K$ a slow algebraic decay of $f_t$ is
seen.  A representative case is $K=0.1$, whose trapping statistics are
given in Fig.\ 3(a), which illustrates the slow decay for very
long times $t\sim 10^7$.  Also given in Fig.\ 3 are $P_t$, $C_\tau$,
and $\alpha\equiv-d\log C_\tau/d\log\tau$ (thus locally
$C_\tau\sim \tau^{-\alpha}$).  This last plot shows the power at which
$C_\tau$ decays varying between about $\fract1/4$ and $\fract3/2$.
\par\topinsert\figinsert{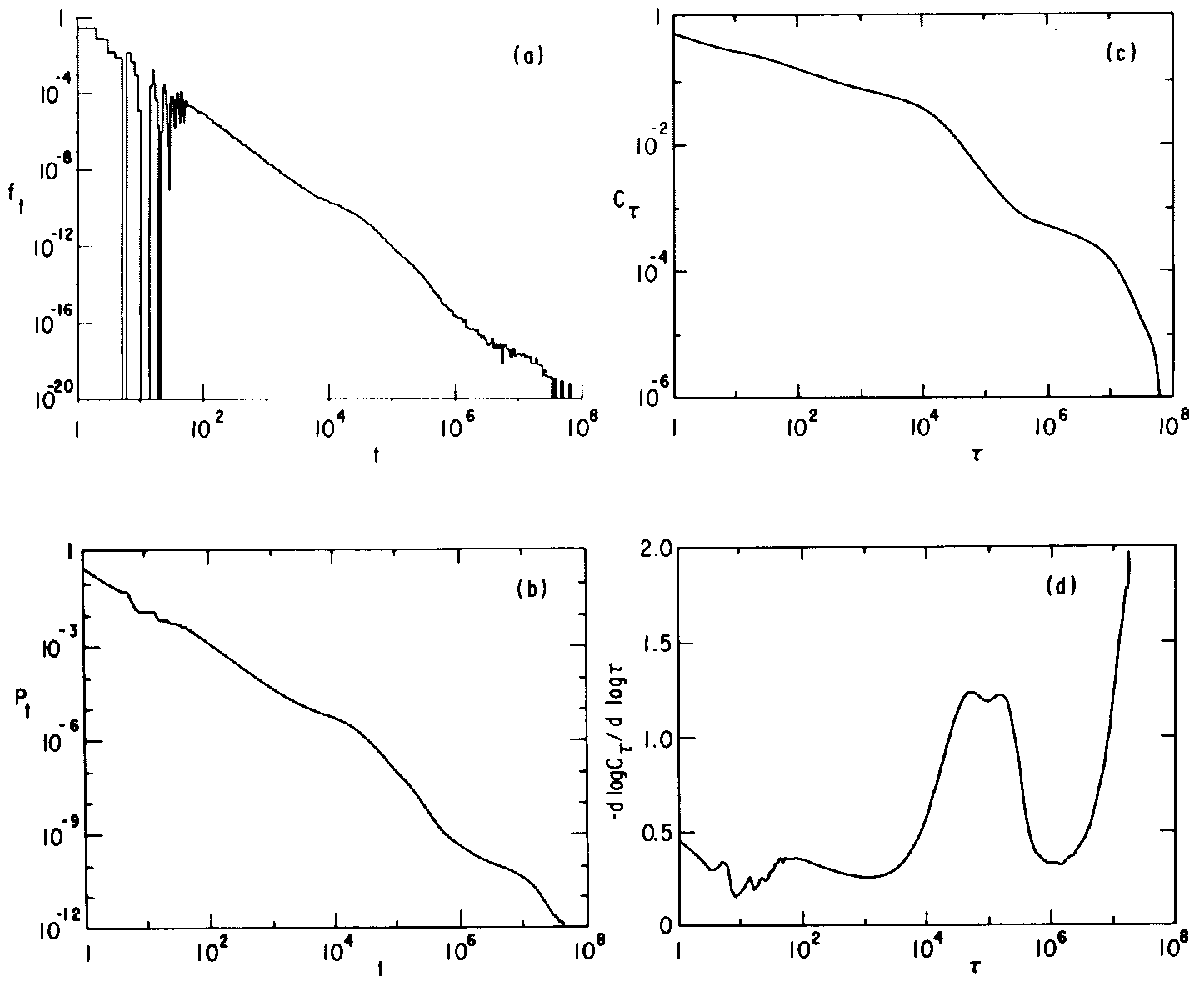}{120}
{\baselineskip 1\vu\noindent
Fig.\ 3.  (a) The trapping statistics $f_t$ for $K=0.1$.  (b), (c),
and (d) show $P_t$, $C_\tau$, and $d\log C_\tau/d\log \tau$.\par}
\vskip 5truemm plus 1truemm\endinsert

A glance at Fig.\ 2 shows the origin of this behavior.  The central
island is surrounded by a chain of sixth-order islands.  Around each of
these islands are several other sets of islands.  This picture repeats
itself at deeper and deeper levels.  A particle which manages to
penetrate into this maze can get stuck in it for a long time.

For $\tau\lsapprox10^4$, Fig.\ 3(d) gives
$\alpha\approx\fract1/4$.  Correspondingly we have $P_t\sim t^{-p}$ where
$p=1+\alpha\approx 5/4$.  This is close to the asymptotic ($t\to\infty$)
result found by Chirikov and Shepelyansky (1981) for the whisker map, in which
$\ave p\approx 1.45$.
However, in our case, $\alpha$ shows some strong variations beyond $\tau
\approx10^4$ where $C_\tau$ ``steps down'' (e.g., between $10^4$ and
$3\times10^5$).  This means that the asymptotic form of $C_\tau$ is
very difficult to determine numerically.

The diffusion coefficient $D$ is given by the summation of $C$ and is
approximately $6400$.  The error in this estimate of $D$ depends on the
asymptotic form for $C_\tau$.  On the basis of the numerical results,
we cannot rule out
the possibility that as $\tau\to\infty$,
$C_\tau$ decays with $\alpha\le1$.  In that case, $D$ would be
infinite!

If $D$ is indeed infinite, we would wish to know how a group
of particles spreads with time.  We again consider
the drunkard's walk based on $\qstar$ which was introduced earlier.
The second moment of $r$ is related to the correlation
function by
$$S_t\equiv\ave{(r_t-r_0)^2}=tC_0+2\sum_{\tau=1}^t(t-\tau)C_\tau.$$
This is plotted in Fig.\ 4(a), using the data of Fig.\ 3.
For $t\lsapprox10^4$, $S_t$ grows somewhat faster than $t^{3/2}$ (see
Fig.\ 4(b)) and even until $t\approx10^7$, $S_t$ is growing
significantly faster than linearly.  Beyond $10^7$, the numerical data
shows a convergence to a linear rate; but this is merely because no
segments longer than about $6\times10^7$ were observed.  For
$t\to\infty$, $S_t$ grows as $t^{2-\alpha}$, assuming that the exponent
$\alpha$ at which $C_\tau$ decays asymptotically is less than
$1$.  If the diffusion coefficient is estimated from $D_t=\half S_t/t$,
then $D_t$ grows with $t$ as shown in Fig.\ 4(c).
\par\topinsert\figinsert{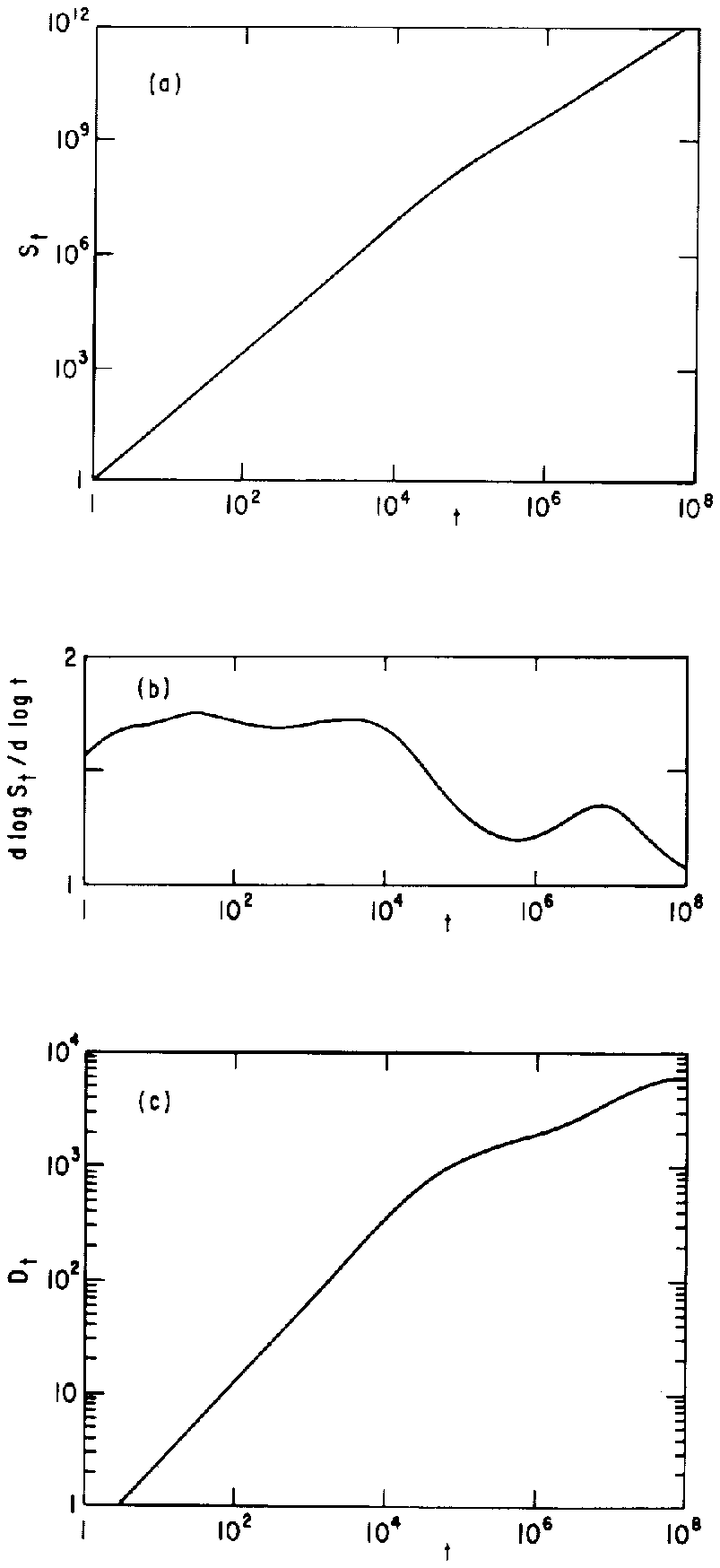}{141}
{\baselineskip 1\vu\noindent
Fig.\ 4.  (a) The variance $S_t$ for the case given in Fig.\
3.  (b) and (c) show $d\log S_t/d\log t$ and
$D_t=\half S_t/t$.\par}\vskip 5truemm plus 1truemm\endinsert

\section Discussion

We can apply these results to the determination of the correlation
function of a general mapping.  Suppose the correlation function is
defined by
$$\Cscr(\tau)=\ave{h(\vec x(t))h(\vec x(t+\tau))}_t,$$
\def\cisl{\Cscr_{\subrm{is}}(\tau)}%
where $h$ is some smooth function of the position in phase space $\vec
x$.  Then the contribution of an
island located at $\vec x_0$ to $\Cscr(\tau)$ is (Karney, 1983)
$$\cisl=h^2(\vec x_0)(B/A)C_\tau,$$
where $A$ is the total area of the stochastic component of the general mapping
and $B$ is the portion of the that area which is near the island.
(More precisely, when we regard the $L\times L$ square of
$\qstar$ as being a magnification of a small area of the general
mapping, then $B$ measures the area of the stochastic component within
this small area.)  Similar relations connect
$D$ and $S_t$ and the corresponding quantities for the general
mapping. 

In the case of accelerator mode in the standard mapping,
$K$ is related to the parameter $k$ by $k^2=(2\pi n)^2+16 K$ where $n$
is an integer.  For $K=0.1$, we take $6400$ as a lower bound for $D$.
We find that the contribution to the diffusion coefficient is
increased over its quasi-linear value by a factor of at least
$360/n^2$.
Thus for $n=1$ or $k\approx 6.41$, the islands completely dominate the
diffusion.  The first-order accelerator modes continue to have such a
large effect at least until $k\approx 100$.
If $D$ is in fact infinite, even arbitrarily small accelerator
modes will eventually dominate the motion and
Fig.\ 4 can be used to estimate
the time at which the accelerator modes become important.

In summary, small islands within the stochastic sea
lead to correlations in the stochastic orbits for extremely long times.
When the islands are accelerator modes, this may cause the particles to
behave non-diffusively, i.e.,
the mean squared
displacement of the particles may increase faster than linearly with time.

In order to provide a definitive answer to this question, the
asymptotic behavior of $C_\tau$ must be determined.  Because the
asymptotic regime starts at such a large $\tau$ (greater than
$10^7$), it appears its properties cannot be studied by
the numerical method used in this paper.  What is needed is a better
analytical understanding of the behavior of trajectories close to the
border between integrability and stochasticity.  Chirikov (1982) has
made some useful steps in this direction.

\section Acknowledgments

This work was supported by the U.S. Department of Energy under Contract
DE--AC02--76--CHO3073.  I first became interested in this problem at
the previous workshop on this subject held in Kyoto in November 1981.
I began working on it while at the Institute of Plasma
Physics, Nagoya University, participating in the U.S.-Japan Fusion
Cooperation Program.  Some of the work was also carried out while at the
Aspen Center for Physics.  I would like to thank J. M. Greene, R. S.
MacKay, and F. Vivaldi for stimulating discussions.  B. V. Chirikov
also provided some enlightening comments.

\section References

\def\Ref. #1\cr{\noindent\hangindent 1.5\vu \hangafter 1 #1\par}%
\def\rule{\null\vrule width 3\vu height 2.4pt depth -2pt\kern 1pt }%
\Ref. S. R. Channon, and J. L. Lebowitz. 1980.
``Numerical Experiments in Stochasticity and Homoclinic Oscillation''
in {\it Nonlinear Dynamics}, Annals of the New York Academy of Sciences
{\bf 357}: 108--118.\cr
\Ref. B. V. Chirikov. 1979.
``A Universal Instability of Many-Dimensional Oscillator Systems.''
{\it Phys.\ Rept.} {\bf 52}: 263--379.\cr
\Ref. \rule. 1982.
``Chaotic Dynamics in Hamiltonian Systems with Divided Phase Space.''
in {\it 7th Sitges Conference on Dynamical Systems and Chaos}, Barcelona.
(Institute of Nuclear Physics Preprint INP 82--132, Novosibirsk.)\cr
\Ref.  \rule, and D. L. Shepelyansky. 1981.
``Statistics of Poincar\'e Recurrences and the Structure of the
Stochastic Layer of a Nonlinear Resonance'' (in Russian) in
{\it 9th International Conference on Nonlinear Oscillations}, Kiev.
(Institute of Nuclear Physics Preprint INP 81--69, Novosibirsk;
English translation: Princeton Plasma Physics Laboratory Report
PPPL--TRANS--133, 1983.)\cr
\Ref. C. F. F. Karney. 1983.
``Long-Time Correlations in the Stochastic Regime.''
{\it Physica} {\bf 8D}:  360--380.\cr
\Ref. \rule, A. B. Rechester, and R. B. White. 1982.
``Effect of Noise on the Standard Mapping.''
{\it Physica} {\bf 4D}: 425--438.\cr

\bye